\documentclass[prx,superscriptaddress,amsmath,amssymb,twocolumn,nofootinbib]{revtex4-2}
\usepackage{graphicx,bm,amsmath}
\usepackage[usenames,dvipsnames]{xcolor}
\usepackage[colorlinks,bookmarks=false,citecolor=NavyBlue,linkcolor=Red,urlcolor=blue,
]{hyperref}
\usepackage[percent]{overpic}
\usepackage[bbgreekl]{mathbbol}
\usepackage{xcolor}
\usepackage[normalem]{ulem}
\usepackage{algorithm}
\usepackage{braket}
\usepackage{algpseudocode}
\usepackage{comment}
\usepackage{amsthm}
\usepackage{enumerate}
\usepackage{lipsum}
\def\doi{http://dx.doi.org/}
\newcommand{\be}{\begin{equation}}
\newcommand{\ee}{\end{equation}}
\newcommand{\bec}{\begin{equation*}}
\newcommand{\eec}{\end{equation*}}
\newcommand{\bea}{\begin{eqnarray}}
\newcommand{\eea}{\end{eqnarray}}

\usepackage{color}

\newcommand{\titleinfo}{Scalable Quantum Monte Carlo for Shannon-Rényi Entropy: An Experiment-visitable Probe of Quantum Phases and Criticality}

\begin{document}
\title{\titleinfo}
\author{Dong-Xu Liu}
\affiliation{Department of Physics, School of Science and Research Center for Industries of the Future, Westlake University, Hangzhou 310030, China}
\affiliation{Institute of Natural Sciences, Westlake Institute for Advanced Study, Hangzhou 310024, China}

\author{Zhe Wang}
\email{zhewang@westlake.edu.cn}
\affiliation{Hangzhou International Innovation Institute, Beihang University, Hangzhou 311115, China}
\affiliation{Department of Physics, School of Science and Research Center for Industries of the Future, Westlake University, Hangzhou 310030, China}
\affiliation{Institute of Natural Sciences, Westlake Institute for Advanced Study, Hangzhou 310024, China}

\author{Zenan Liu}
\affiliation{Department of Physics, South China University of Technology, Guangzhou 510640, China}

\author{Yi-Ming Ding}
\affiliation{Department of Physics, School of Science and Research Center for Industries of the Future, Westlake University, Hangzhou 310030, China}
\affiliation{Institute of Natural Sciences, Westlake Institute for Advanced Study, Hangzhou 310024, China}

\author{Zheng Yan}
\email{zhengyan@westlake.edu.cn}
\affiliation{Department of Physics, School of Science and Research Center for Industries of the Future, Westlake University, Hangzhou 310030, China}
\affiliation{Institute of Natural Sciences, Westlake Institute for Advanced Study, Hangzhou 310024, China}

\begin{abstract}
The Shannon-R\'enyi (SR) entropy characterizes the probability distribution of many-body wave functions in a chosen basis and is directly accessible from single-shot measurement snapshots in cold-atom experiments, yet its theoretical evaluation in interacting systems carries exponential cost. Here we construct a field-theoretic spacetime manifold for the SR entropy in which $q$ replicas are glued at the imaginary-time boundaries, recasting it as a ratio of generalized partition functions amenable to large-scale quantum Monte Carlo simulation. On this basis we develop a reweight-annealing framework with two complementary strategies: a virtual-parameter scheme interpolating between unglued and fully glued limits via binomial-weighted partial gluing, and a real-parameter scheme using a physical coupling constant as the annealing variable within stochastic series expansion. Both reduce the computational cost from exponential to polynomial. Most importantly, we establish for the first time a finite-size scaling theory for the SR entropy near a quantum critical point, $S_q=a(g)L^2+S_q^0\left((g-g_c)L^{1/\nu}\right)$; data collapse of its first derivative yields the correlation-length exponent $\nu$ for both the O(3) and Ising universality classes, realized in the spin-$\frac{1}{2}$ columnar dimerized Heisenberg model and the bilayer Heisenberg-Ising model. The same simulations extract the Goldstone-mode count $N_G=2$ and the $Z_2$ ground-state degeneracy $\ln 2$ from subleading terms. These results elevate the SR entropy to a standalone, order-parameter-free diagnostic of quantum criticality with the same scaling status as the entanglement entropy, and close the loop between theory and quantum simulation by enabling quantitative benchmarking of experimental snapshots from state-of-the-art quantum simulators.
\end{abstract}

\maketitle
\section{Introduction}
The Shannon-R\'enyi (SR) participation entropy provides a powerful diagnostic 
of the structure of quantum many-body wavefunctions in a chosen measured basis \cite{Santos2010Onset,Zaletel2011Logarithmic,Stephan2013Entanglement,Stephan2014Geometric,Luitz2015Many}. Given a quantum state described by the density matrix $\rho$, its diagonal 
elements define a probability distribution $p_{s} = \rho_{ss} = \bra{s}\rho\ket{s}$, with $\sum_s p_s = 1$. For a pure ground state $\ket{\Psi_0} = \sum_s \psi_s \ket{s}$, this reduces to $p_s = |\psi_s|^2$. The $q$-th Shannon-R\'enyi entropy is then
\begin{equation}\label{eq:SR_entropy}
    S_q = \frac{1}{1-q} \ln \sum_{s} (\rho_{ss})^{q} ,
\end{equation}
where $\rho_{ss}$ denotes the diagonal part of the density matrix in the chosen basis. The Shannon limit is recovered as $q \to 1$, yielding 
$S_1 = -\sum_s \rho_{ss} \ln \rho_{ss}$ which is named Shannon participation entropy. In the basis that diagonalizes the density matrix, the Shannon participation entropy reduces to the von Neumann entropy. In the $q \to \infty$ limit, it reduces to $S_\infty = -\ln(\max_{s}\left\{ \rho_{ss}\right\} )$\cite{Luitz2014Shannon,Luitz2014Universal}, which depends solely on the largest diagonal element and is computationally the simplest. In contrast to the R\'enyi entanglement entropy which is defined via the reduced density matrix $\rho_A$ of a spatial subregion as $S_q^{\rm ent} = \frac{1}{1-q} \ln \operatorname{tr}(\rho_A^q)$ and is manifestly basis-independent, the SR entropy involves only the diagonal elements of the full density matrix and is inherently basis-dependent, requiring no spatial bipartition \cite{Alcaraz2013Universal,Alcaraz2014Universal,Leone2022Stabilizer,Trino2026Stabilizer}. 

For generic many-body states with ${N}$ lattice sites, the SR entropy typically obeys a volume-law scaling at leading order \cite{Stephan2009Shannon}: 
\begin{equation}\label{eq:scaling}
    S_q(N) = a_q N + l_{q} \ln(N) + b_{q}
\end{equation}
Here, $a_{q}$ ($q > 1$) is a non-universal coefficient that depends on the microscopic model and on $q$. The nonlinear $q$-dependence of $a_{q}$ signals intrinsic multifractality as a generic feature of interacting quantum many-body systems \cite{Evers2000Fluctuations,Evers2008Anderson,Atas2012Multifractality}. At continuous phase transition points described by conformal field theory, $l_q=0$ and $b_{q}$ takes a universal constant value; for example, at the three-dimensional O(3) critical point, a universal constant term $b_{q}$ appears\cite{Luitz2014Universal}. In the continuous symmetry-breaking phase, the logarithmic subleading term is directly tied to the number of Goldstone modes, distinguishing continuous from discrete symmetry breaking \cite{Atas2012Multifractality,Stephan2014Shannon,Luitz2014Universal,Luitz2017Quantum,Misguich2017Finite,Nicolas2016Quantum,Metlitski2015Entanglement}. In gapped or discrete symmetry-breaking phases, the constant $b_{q}$ may encode the ground-state degeneracy \cite{Stephan2009Shannon,Stephan2010Renyi,Stephan2014Shannon}. These properties establish the SR entropy as a general-purpose tool for classifying quantum phases without prior knowledge of an order parameter, similar to what entanglement entropy does \cite{Nicolas2016Quantum}. However, compared with the challenge in extracting entanglement entropy in experiment, the SR entropy is friendly for measurement in cold-atom experiments \cite{koyluouglu2026measuring}, especially when the snapshots can be easily obtained like in Rydberg arrays.
Naturally, it is important to extend the understanding for this information-theoretic quantity in quantum many-body systems in theory and numeric simultaneously.

However, a fundamental gap remains in the theoretical understanding of the SR entropy. For the entanglement entropy, quantum criticality is known to imprint a universal finite-size scaling structure: close to a critical point $g_c$, after subtracting the non-universal leading contribution, the entropy collapses onto a single-parameter scaling function of $(g-g_c)L^{1/\nu}$ governed by the correlation-length exponent $\nu$ \cite{Fradkin2006Entanglement,Oshikawa2010Boundary,Calabrese2004,Calabrese2009}. This single-parameter scaling law is what elevates the entanglement entropy from a phase classifier to a quantitative diagnostic of criticality, allowing critical points and exponents to be extracted without reference to any order parameter. Whether the SR entropy, which probes the full rather than the reduced density matrix, is basis-dependent, and requires no bipartition, obeys the same universal scaling behavior near a quantum critical point has remained unknown. The subleading terms in Eq.~\eqref{eq:scaling} are understood only at the critical point or deep inside a phase; there is no scaling theory describing how the SR entropy behaves as the critical point is approached, and no demonstration that it shares the scaling structure of the entanglement entropy. Closing this gap is essential: it would promote the SR entropy to a diagnostic of quantum criticality on the same theoretical footing as the entanglement entropy, but one that is directly accessible to experiment.

Traditional methods for evaluating the SR entropy include exact diagonalization\cite{Stephan2011Phase}, restricted to small system sizes by the exponential growth of the Hilbert space, and density matrix renormalization group (DMRG) techniques, which compute the SR entropy directly from its definition by constructing the diagonal elements of the density matrix in the chosen computational basis, subject to the limitation of bond dimension \cite{Schuch2008Entropy,Tagliacozzo2008Scaling,Pollmann2009Theory}. Within the quantum Monte Carlo (QMC) framework\cite{Sandvik1997Quantum,Syljusen2002Quantum,syljusen2003directed,Evertz2003The,Fabien2005Generalized,yan2019sweeping,yan2022global}, previous attempts evaluate the SR entropy by directly counting a histogram over configuration space; this is equivalent to sampling the density matrix $\rho = e^{-\beta H}/Z$ and subsequently computing $\sum_s \rho_{ss}^{\,q}$ from the sampled configurations \cite{Luitz2014Universal,Mao2025Sampling,Mao2026Detecting}. However, when the ground-state wavefunction is delocalized over many basis configurations, which is precisely the regime near a critical point where the SR entropy is large and physically most informative, the number of non-negligible $\rho_{ss}$ entries scales exponentially with system size. Consequently, the computational cost of directly accumulating the diagonal sum grows exponentially, and the statistical error in the QMC estimate becomes uncontrollable. Existing QMC-based algorithms for the SR entropy are therefore prohibitively expensive and essentially brute-force in character, restricting their applicability to small systems or to special parameter regimes where the wavefunction is already strongly localized.

A promising route to overcoming this exponential bottleneck is offered by the QMC algorithms widely used for the R\'enyi entanglement entropy \cite{Helmes2014Entanglement,luitz2014improving,Demidio2020Entanglement,Zhao2022Scaling,wang2025bipartite,Ding2025Tracking}, whose efficiency rests on representing the entanglement path integral on a sheeted spacetime manifold that is easy to simulate \cite{1994NuPhB.424..443H,Calabrese2004,Calabrese2009,Fradkin2006Entanglement,Oshikawa2010Boundary,Berthiere2016Boundary,Angel-Ramelli2019Entanglement}. 
No analogous spacetime-manifold representation has so far existed for the SR entropy, and this absence has strongly hindered the development of QMC methods for this quantity: while the SR entropy is readily accessible in experiments, theoretical computation has remained confined to tomography of sampled configurations \cite{Luitz2014Shannon} with exponential complexity, a clear departure from the polynomial-complexity advantage that QMC normally offers for many-body problems.

In this work, we resolve both the conceptual and the computational gaps identified above. Our central conceptual result is the first finite-size scaling theory for the SR entropy in the vicinity of a quantum critical point: we propose the scaling form $S_q = a(g)L^2 + S_q^0 \!\left((g-g_c)L^{1/\nu}\right)$, which requires neither spatial bipartition nor partial trace, and we demonstrate that data collapse of its parametric derivative with respect to $g$ directly yields the correlation-length exponent $\nu$. This promotes the SR entropy from a phase classifier to a quantitative diagnostic of quantum criticality. On the computational side, we design a field-theoretic replica spacetime representation for the SR quantities and develop a unified stochastic series expansion (SSE) QMC framework to evaluate the SR entropy and its derivatives in a numerically exact and scalable manner, reducing the computational cost from an exponential dependence on system size to a polynomial one. We verify the scaling form $S_q = a(g)L^2 + S_q^0\!\left((g-g_c)L^{1/\nu}\right)$ on spin models belonging to distinct universality classes, namely the columnar dimerized Heisenberg model and the bilayer Heisenberg-Ising model, and show that the SR entropy simultaneously reveals the nature of symmetry breaking, counts Goldstone modes, identifies the universality class, and extracts critical exponents, all without any prior knowledge of an order parameter. By providing a scalable and exact numerical toolbox together with its critical scaling theory, this work not only unlocks the diagnostic power of the SR entropy for complex quantum phases but also paves the way for a direct, quantitative comparison between theoretical predictions and ongoing quantum-simulator experiments.

\section{Methods}
\subsection{Spacetime manifold of SR entropy}
Consider a quantum many-body system described by the Gibbs state $\rho = e^{-\beta H} / Z$ at inverse temperature $\beta$, and the partition function $Z = \operatorname{tr} e^{-\beta H}$. Working in a complete orthonormal measured basis $\{\ket{s}\}$ (e.g., the $S^z$ basis), the diagonal elements of the density matrix are $\rho_{ss} = \bra{s}\rho\ket{s} = Z^{-1} \bra{s}e^{-\beta H}\ket{s}$. The ${q}$-th Shannon-R\'enyi entropy $S_{q}$ is defined through the sum over diagonal weights:
\begin{equation}\label{eq:Zalpha}
    \mathcal{R}_{q} \equiv \sum_s \rho_{ss}^{q}
    = \frac{1}{Z^{q}} \sum_s \bigl[ \bra{s} e^{-\beta H} \ket{s} \bigr]^{q},
\end{equation}
which yields
\begin{equation}
    S_{q} = \frac{1}{1-q} \ln \mathcal{R}_{q}.
\end{equation}
Eq.~\eqref{eq:Zalpha} involves the $q$-th power of individual diagonal matrix elements of $\rho$, not the $q$-th power of the reduced density matrix as in entanglement entropy ($\operatorname{tr}\rho_A^{q}$). The latter requires a non-trivial partial trace over spatial degrees of freedom, whereas Eq.~\eqref{eq:Zalpha} treats each basis-state diagonal element as a scalar quantity.

\begin{figure}[ht!]
\centering
\includegraphics[width=1.0\linewidth]{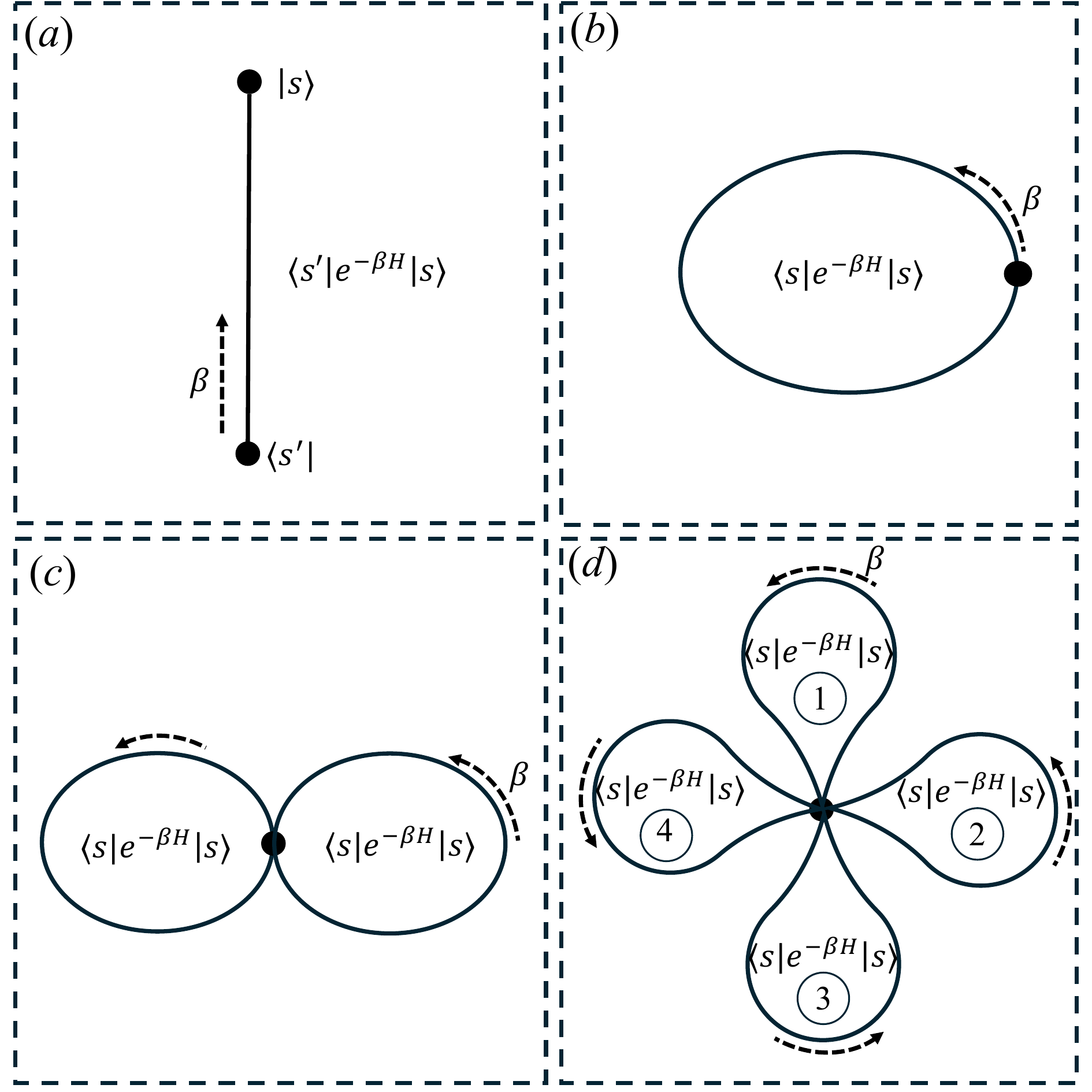}
\caption{Manifold representation of the density-matrix element $[\langle s\vert e^{-\beta H}\vert s \rangle]^{q}$ with gluing boundary conditions along the imaginary-time direction. 
(a) Density-matrix element 
$\langle s^{\prime} \vert e^{-\beta H} \vert s \rangle$. (b) Diagonal element $\langle s \vert e^{-\beta H} \vert s\rangle$. Black solid points denote basis states; the loop forms because $\vert s^{\prime} \rangle = \vert s \rangle$. (c) Manifold representing $[\langle s \vert e^{-\beta H} \vert s \rangle]^{2}$. The two replicas share the same state $\vert s \rangle$ at the imaginary-time boundaries $\tau=0$ and $\tau=\beta$. (d) Manifold representation of the factor $[\langle s \vert e^{-\beta H}\vert s \rangle]^{4}$ used in the fourth-order SR entropy. The four replicas share the same state $\vert s \rangle$ under the gluing imaginary-time boundary condition at $\tau=0$ and $\tau=\beta$ in each replica.}
\label{fig:manifold}
\end{figure}

As mentioned above, the challenge for applying large-scale QMC in SR entropy is that the spacetime manifold of path-integral is unknown so far. In this section, we want to address this problem. Considering an arbitrary path integral as the Fig.~\ref{fig:manifold} (a) shows, a state at a fixed time is represented as a point and the path of imaginary time is a line, where the length of the line is the distance in imaginary time $\beta$ \cite{Mao2025Sampling,Ding2026mixed,ding2026quantummontecarloage}. The figure actually displays an imaginary time evolution between the states $s$ and $s'$, i.e., $\langle s'| e^{-\beta H} |s \rangle$. If we connect the two endpoints together, the boundary condition of imaginary time has been changed to enforce a diagonal path-integral $\langle s| e^{-\beta H} |s \rangle$, as shown in the Fig.~\ref{fig:manifold} (b).
In this way, the spacetime manifold of $[\langle s| e^{-\beta H} |s \rangle]^2$ can be represented as two connected loops, as the Fig.~\ref{fig:manifold} (c) shows. The two replicas share the same basis state $\vert s \rangle$ at the imaginary-time boundaries $\tau=0$ and $\tau=\beta$, i.e., the boundary spin configuration is identical on every lattice site, so that the $q=2$ SR manifold has been constructed. Following this idea, the $q$-th SR manifold is constructed by $q$ loops glued at the imaginary-time boundaries, as the example $q=4$ shows in the Fig.~\ref{fig:manifold} (d). Consequently, we use $Z^{(q)}_G$ to represent the generalized partition function of $q$ glued loops, thus the Eq.~\eqref{eq:Zalpha} can be rewritten as:
\begin{equation}\label{eq:factorization}
    \mathcal{R}_{q} =  \frac{Z^{(q)}_G}{Z^{q}}.
\end{equation}
The SR entropy is now expressed as a ratio of generalized partition functions, a form directly amenable to QMC methods~\cite{Demidio2020Entanglement,Demidio2024Entanglement,wang2025bipartite,Zhao2022Scaling}.

\subsection{Reweighting virtual parameter: Calculation of SR entropy}
Construct an interpolating partition function $\mathcal{Z}(\lambda)$ parametrized by a virtual parameter $\lambda \in [0,1]$ that matches the two physical limits at the endpoints:
\begin{align}
    \mathcal{Z}(0) &= Z^{q}      &&\text{(completely unglued)}, \label{eq:Zlambda_a} \\
    \mathcal{Z}(1) &= Z^{(q)}_G &&\text{(completely glued)}.   \label{eq:Zlambda_b}
\end{align}
Once the ratio $\mathcal{Z}(1)/\mathcal{Z}(0)$ is accessible, the target entropy follows as
\begin{equation}
    S_{q} = \frac{1}{1-q} \ln \frac{\mathcal{Z}(1)}{\mathcal{Z}(0)}.
\end{equation}

The parameter $\lambda$ plays the role of a ``gluing strength.'' At $\lambda = 0$ the $q$ replicas are completely independent with uncorrelated boundary configurations; at $\lambda = 1$ they are forced to identical boundary configurations; intermediate values $0 < \lambda < 1$ describe partial gluing, where the boundary spin configurations of the $q$ replicas are forced to agree only on a subset $B$ of lattice sites (see below).

\textbf{Binomial weight construction:} For clarity, we first present the construction for $q=2$ and a spin model, where each lattice site has a binary gluing status. Each boundary configuration is a product state $\vert s \rangle = \vert s_1 s_2 \cdots s_N \rangle$ specified by the $N$ spin variables $s_i$ on the $N$ lattice sites. Let $\mathcal{S} = \{1, 2, \dots, N\}$ denote the set of all $N$ lattice sites, and let $B \subseteq \mathcal{S}$ be the subset of sites on which the two replicas are forced to agree at the imaginary-time boundaries: for every site $i \in B$ the boundary spin $s_i$ is identical in both replicas, whereas for every site $i \in \mathcal{S} \setminus B$ the two boundary spins are independent. The interpolating partition function is defined as a binomial-weighted mixture over\cite{Humeniuk2012Quantum,Demidio2020Entanglement,Demidio2024Entanglement}:
\begin{equation}\label{eq:Zbinomial}
    \mathcal{Z}(\lambda) =
    \sum_{B \subseteq \mathcal{S}}
    \lambda^{|B|} (1-\lambda)^{N - |B|} \; \mathcal{Z}(B),
\end{equation}
where $\mathcal{Z}(B)$ is the constrained partition function in which the two replicas share identical boundary spins on the sites of $B$, while the boundary spins on $\mathcal{S} \setminus B$ are free and independent in each replica. $|B|$ is the number of sites in the area $B$. The generalization to $q>2$ follows by replacing the binomial weight with a sum over all partitions of the $q$ replicas, where each partition specifies which replicas share the same boundary configuration on a given site.

The statistical interpretation is natural: each basis-state index independently participates in gluing with probability $\lambda$. Out of $N$ independent Bernoulli distribution, the probability that exactly $|B|$ indices are glued is $\binom{N}{|B|} \lambda^{|B|} (1-\lambda)^{N - |B|}$, precisely the weight factor in Eq.~\eqref{eq:Zbinomial}.

\textbf{Increment along $\lambda$}: The distributions of $\mathcal{Z}(1)$ and $\mathcal{Z}(0)$ have vanishing overlap. Subdividing the virtual-parameter interval $[0,1]$ into $M$ narrow sub-windows $[\lambda_{k-1}, \lambda_k]$ ensures that the weight distributions at adjacent $\lambda$ values overlap strongly, and their weight ratio differs only by $\mathcal{O}(\Delta\lambda)$.

The target ratio is expanded as a telescoping product of $M$ adjacent ratios\cite{Neal2001,Humeniuk2012Quantum,Zhao2022Scaling,dai2025residual,Ding2025Tracking,wang2025,Ding2025Evaluating}:
\begin{equation}\label{eq:product}
    \frac{\mathcal{Z}(1)}{\mathcal{Z}(0)}
    = \frac{\mathcal{Z}(\lambda_1)}{\mathcal{Z}(\lambda_0)}
      \frac{\mathcal{Z}(\lambda_2)}{\mathcal{Z}(\lambda_1)} \cdots
      \frac{\mathcal{Z}(\lambda_M)}{\mathcal{Z}(\lambda_{M-1})},
\end{equation}
with $\lambda_0 = 0$ and $\lambda_M = 1$.  Equation~\eqref{eq:product}
is the core of the reweight-annealing algorithm~\cite{Ding2024Reweight,wang2025bipartite,Wang2026Addressing}: it converts the ratio of two quasi-orthogonal distributions into a product of $M$ ratios of adjacent, strongly overlapping distributions, thereby resolving the numerical catastrophe at its root.

We now derive the computational formula for each sub-window ratio. Starting from the definition Eq.~\eqref{eq:Zbinomial}:
\begin{align}
    \frac{\mathcal{Z}(\lambda_k)}{\mathcal{Z}(\lambda_{k-1})}
    &= \frac{ \sum_{B} \lambda_k^{|B|} (1-\lambda_k)^{N - |B|} \mathcal{Z}(B) }
            { \sum_{B} \lambda_{k-1}^{|B|} (1-\lambda_{k-1})^{N - |B|} \mathcal{Z}(B) }
       \nonumber \\[6pt]
    &= \Bigl\langle
       \Bigl( \frac{\lambda_k}{\lambda_{k-1}} \Bigr)^{|B|}
       \Bigl( \frac{1-\lambda_k}{1-\lambda_{k-1}} \Bigr)^{N - |B|}
       \Bigr\rangle_{\lambda_{k-1}}.
    \label{eq:Zratio}
\end{align}
In the second equality we multiply and divide the numerator by $\lambda_{k-1}^{|B|}$ and $(1-\lambda_{k-1})^{N - |B|}$, recognizing that $\frac{1}{\mathcal{Z}(\lambda_{k-1})} \lambda_{k-1}^{|B|}(1-\lambda_{k-1})^{N - |B|}\mathcal{Z}(B)$ is precisely the equilibrium probability of configuration~$B$ at $\lambda_{k-1}$, thereby converting the sum into an ensemble average. Here $\langle \cdot \rangle_{\lambda_{k-1}}$ denotes the ensemble average with respect to the equilibrium distribution at $\lambda = \lambda_{k-1}$, and $|B|$ is the number of glued lattice sites at the imaginary-time boundaries. 

\medskip\noindent
\textbf{Metropolis Acceptance Probabilities}: During Monte Carlo sampling at a fixed $\lambda$, the Metropolis updates are used that toggle the gluing status of a lattice site.
Before proposing gluing, one must verify that the boundary configurations of the two replicas at the chosen site are identical.  If they differ, the move is rejected immediately regardless of $\lambda$.  When they match:
\begin{equation}\label{eq:accept01}
    p_{\rm accept}(|B| \to |B|+1)
    = \min\!\Bigl( 1, \; \frac{\lambda}{1-\lambda} \Bigr).
\end{equation}

\medskip\noindent
Similarly, the ungluing process can be designed:
\begin{equation}\label{eq:accept10}
    p_{\rm accept}(|B| \to |B|-1)
    = \min\!\Bigl( 1, \; \frac{1-\lambda}{\lambda} \Bigr).
\end{equation}
The similar operation can be found in the QMC simulation of entanglement entropy \cite{Demidio2020Entanglement}.

\subsection{Reweighting with a Real Parameter}
The virtual-parameter construction introduced above requires building an artificial binomial interpolation between the glued and unglued replica manifolds. While conceptually transparent, this approach incurs substantial computational overhead because all the intermediate ratios have no physical meaning. An alternative strategy, employed in recent QMC studies of the R\'enyi entanglement and R\'enyi negativity, is to use a real physical parameter, such as the coupling constant $J_1$ as the reweighting variable\cite{wang2025bipartite,Ding2025Tracking,Wang2026Addressing}. This eliminates the need for an auxiliary interpolation and directly connects the SR entropy to tuning the Hamiltonian. All the intermediate values are also the related physical quantities.

Consider the columnar dimerized Heisenberg Hamiltonian in Eq.~\eqref{eq:HJ1J2} with fixed $J_2$ and variable $J_1$. The SR entropy at coupling $J_1$ is
\begin{equation}
    S_q(J_1) = \frac{1}{1-q} \ln\!\left[ \frac{Z^{(q)}_{G}(J_1)}{Z^{q}(J_1)} \right],
\label{eq:Sq_J1}
\end{equation}
To evaluate $S_q$ at a target coupling $J^{\prime}_{1}$, we start the reweight-annealing from a reference point $J_1$, and divide the interval $[J_1, J^{\prime}_{1}]$ into $M$ sub-intervals with grid points $J_1 = J_{1,0} < J_{1,1} < \cdots < J_{1,M} = J^{\prime}_{1}$, writing the ratios as telescoping products:
\begin{align}
    \frac{Z(J^{\prime}_1)}{Z(J_{1})} &= \prod_{k=1}^{M} \frac{Z(J_{1,k})}{Z(J_{1,k-1})},
    \label{eq:prodZ} \\[4pt]
    \frac{Z^{(q)}_{G}(J^{\prime}_{1})}{Z^{(q)}_{G}(J_{1})} &= \prod_{k=1}^{M} \frac{Z^{(q)}_{G}(J_{1,k})}{Z^{(q)}_{G}(J_{1,k-1})}.
    \label{eq:prodZD}
\end{align}
The SR entropy then follows as
\begin{equation}
    S_q(J^{\prime}_{1}) = S_q(J_{1}) + \frac{1}{1-q} \left[ \sum_{k=1}^{M} \ln R_{G,k} - q \sum_{k=1}^{M} \ln R_k \right],
\label{eq:Sq_real}
\end{equation}
where $R_k \equiv Z(J_{1,k})/Z(J_{1,k-1})$ and $R_{G,k} \equiv Z^{(q)}_{G}(J_{1,k})/Z^{(q)}_{G}(J_{1,k-1})$.

To evaluate each sub-window ratio, we work directly within the SSE-QMC framework. In SSE, the partition function is a sum over operator-string configurations $C$. Each configuration carries a weight that factorizes by bond type:
\begin{align}
    Z(J_1) &= \sum_C W(C; J_1, J_2), \\[4pt]
    W(C; J_1, J_2) &\propto J_1^{\,n_1(C)} J_2^{\,n_2(C)},
\label{eq:W_SSE}
\end{align}
where $n_1(C)$ and $n_2(C)$ count the number of $J_1$ and $J_2$ operators in the operator string with configuration $C$, and the omitted prefactors ($\frac{(-\beta)^{n}}{n!}$ with number of nonidentical operators $n$) are independent of $J_1$. The ratio of partition functions at adjacent $J_1$ values then reduces to a reweighting of the SSE operator counts:
\begin{align}
    \frac{Z(J_{1,k})}{Z(J_{1,k-1})} 
    &= \frac{ \sum_C W(C; J_{1,k})  }{ \sum_C W(C; J_{1,k-1}) } \\
    &= \frac{ \sum_C W(C; J_{1,k-1}) \frac{W(C; J_{1,k})}{W(C; J_{1,k-1})}}{ \sum_C W(C; J_{1,k-1}) } \nonumber \\
    &= \frac{ \sum_C W(C; J_{1,k-1}) \; \bigl(J_{1,k} / J_{1,k-1}\bigr)^{n_1(C)} }
            { \sum_C W(C; J_{1,k-1}) } \nonumber \\
    &= \Bigl\langle \bigl( J_{1,k} / J_{1,k-1} \bigr)^{n_1} \Bigr\rangle_{Z(J_{1,k-1})},
\label{eq:Rk_SSE}
\end{align}

The ratio $R_{G,k}$ for the $q$-power-weighted diagonal sum is evaluated analogously, but with the sampling performed on the $q$-replica manifold subject to gluing boundary conditions at $\tau = 0, \beta$:
\begin{equation}
    R_{G,k} = \Bigl\langle \bigl( J_{1,k} / J_{1,k-1} \bigr)^{n_{1}^{(G)}} \Bigr\rangle_{Z^{(q)}_{G}(J_{1,k-1})},
\label{eq:RDk_SSE}
\end{equation}
where $n_{1}^{(G)}$ is the total number of $J_1$ operators counted on the glued $q$-replica manifold.

By choosing a reasonably dense grid, the weight distributions at adjacent $J_1$ values overlap strongly and each reweighting factor remains $\mathcal{O}(1)$, guaranteeing efficient importance sampling and polynomial computational complexity~\cite{Ding2024Reweight}. The reference point $J_1$ corresponds to an easily solvable point, where $S_q$ can be evaluated analytically or through a simple simulation. Annealing $J_1$ outward to the target value then yields the curve of $S_q(J_{1})$ along $J_1$. More generally, the reference value $S_q(J_1)$ at an arbitrary starting point can be obtained from the virtual-parameter method described above.

\subsection{Derivative of SR entropy measured in QMC}
Having established both annealing strategies, we now turn to the parametric derivatives of the SR entropy, which are the key quantities for extracting critical information. To characterize how the wavefunction delocalization encoded in the SR entropy evolves with Hamiltonian parameters, we compute its derivative with respect to the Hamiltonian
\begin{equation}
    H = J_1 H_1 + J_2 H_2,
\label{eq:HJ1J2}
\end{equation}
where $H_1$ and $H_2$ denote the sums over weak and strong bonds, respectively. The derivative of the SR entropy (DSRE) with respect to $J_1$ is therefore
\begin{equation}
    \frac{d S_q}{d J_1}
    = \frac{1}{1-q} \,
      \frac{d}{d J_1}
      \ln\!\left( \frac{Z^{(q)}_{G}}{Z^{q}} \right)
    = \frac{1}{1-q}
      \left[ \frac{1}{Z^{(q)}_{G}} \frac{d Z^{(q)}_{G}}{d J_1} - \frac{q}{Z} \frac{d Z}{d J_1}
      \right].
\label{eq:DSRE_step1}
\end{equation}

The derivative of the ordinary partition function follows 
\begin{equation}
    \frac{d [\operatorname{tr} (e^{-\beta H})]}{d J_1}
    = -\beta \, \operatorname{tr}( e^{-\beta H} \, \frac{d H}{d J_1}).
\label{eq:op_deriv_simple}
\end{equation}
Hence,
\begin{equation}
    \frac{d \ln Z}{d J_1}
    = \frac 1 Z \frac{d}{d J_1} \operatorname{tr} e^{-\beta H}
    = -\beta \, \Bigl\langle \frac{d H}{d J_1} \Bigr\rangle_Z
\label{eq:dZ}
\end{equation}
where $\langle \cdot \rangle_Z \equiv Z^{-1} \operatorname{tr}(\cdot \; e^{-\beta H})$ denotes the canonical thermal average at inverse temperature $\beta$.

Similarly, for the generalized partition function on the glued manifold, we obtain
\begin{equation}
   \frac{d \ln Z^{(q)}_G}{d J_1}
   =\frac{1}{Z^{(q)}_{G}} \frac{d Z^{(q)}_{G}}{d J_1}
    = -\beta \, q\Bigl\langle \frac{d H}{d J_1} \Bigr\rangle_{Z^{(q)}_{G}}.
\label{eq:dDq}
\end{equation}

Substituting Eqs.~\eqref{eq:dZ} and~\eqref{eq:dDq} into Eq.~\eqref{eq:DSRE_step1} gives the intermediate result
\begin{equation}
        \frac{d S_q}{d J_1}
        = \frac{1}{1-q}
          \left[
              -\beta \, q\Bigl\langle \frac{d H}{d J_1} \Bigr\rangle_{Z^{(q)}_{G}}
            + \beta \, q\Bigl\langle \frac{d H}{d J_1} \Bigr\rangle_Z
          \right]
    .
\label{eq:DSRE_intermediate}
\end{equation}
Here $\langle d H / d J_{1} \rangle_{Z^{(q)}_{G}}$ is the expectation value of $d H/d J_1$ taken with respect to the $q$-th SR ensemble, and $\langle d H / d J_1 \rangle_Z$ is the standard thermal average. Similar spirit for calculating the derivatives of entanglement entropy or negativity has been used in Refs.~\cite{wu2020entanglement,wang2025bipartite,Ding2025Evaluating}.

For the columnar dimerized Heisenberg Hamiltonian~\eqref{eq:HJ1J2}, the derivative is $\partial H/\partial J_1 = H_1$. Using the SSE relation
\begin{equation}
    \langle H_1 \rangle = -\frac{\langle n_1 \rangle}{\beta J_1},
\end{equation}
where $n_1$ is the number of $H_1$ operators in the operator string, we obtain the final compact form
\begin{equation}
        \frac{d S_q}{d J_1}
        = \frac{1}{1-q}
          \left[
              \frac{\langle n^{(G)}_{1} \rangle_{Z^{(q)}_{G}}}{J_1}
            - q \, \frac{\langle n_{1} \rangle_Z}{J_1}
          \right].
\label{eq:DSRE_final}
\end{equation}
Here, $n^{(G)}_1$ denotes the total number of $J_1$ operators counted over all $q$ replicas on the glued manifold $Z^{(q)}_{G}$, whereas in the conventional ensemble $Z$ it refers to the count in a single replica. This convention is adopted throughout the paper, and it accounts for the explicit factor of $q$ multiplying the second term. Eq.~\eqref{eq:DSRE_intermediate} and~\eqref{eq:DSRE_final} express the DSRE as a difference of sub-Hamiltonian expectation values between two spacetime manifolds, one evaluated in the $q$-weighted diagonal ensemble and the other in the conventional canonical ensemble.

\subsection{Model}
\begin{figure}[htp]
\centering
\includegraphics[width=1.0\columnwidth]{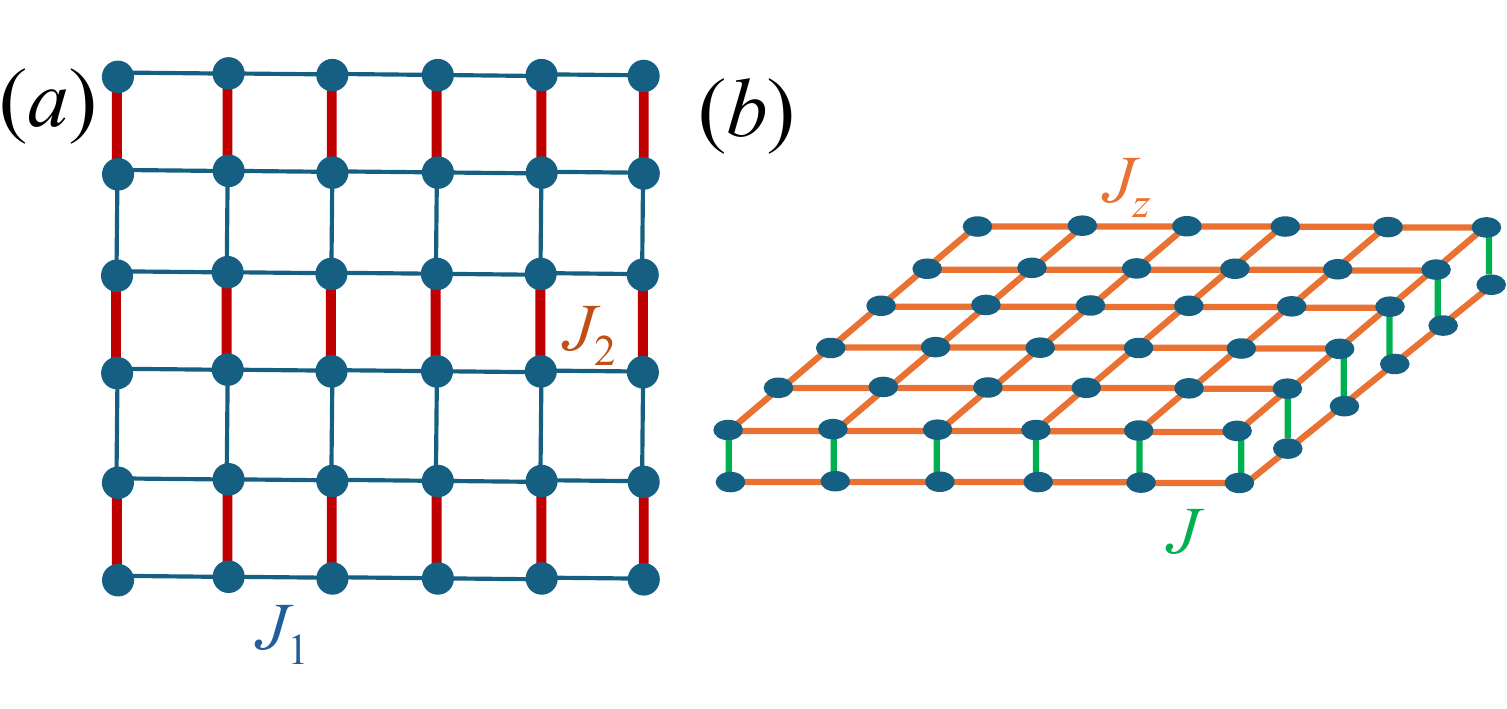}
\caption{Panel (a) illustrates the columnar dimerized Heisenberg model on the square lattice. Different colors indicate the strong-bond and weak-bond sites on the lattice. Panel (b) illustrates the bilayer lattice with a ferromagnetic Ising interaction $J_{z}$ within each layer and a Heisenberg interaction $J$ between the two layers.}
\label{fig:model}
\end{figure}

With the algorithm now fully specified, we introduce the two models used to benchmark the framework. We consider the spin-$\frac{1}{2}$ antiferromagnetic (AF) Heisenberg model with strong and weak bonds on the square lattice (referred to as the columnar dimerized Heisenberg model), defined by the Hamiltonian\cite{Chandra1988Possible,Oitmaa1996Series,Matsumoto2001Ground,Sandvik2010Computational,Zhu2022Exotic,zhu2026bipartite}:
\begin{equation}
    H=J_{1} \sum_{\langle i,j\rangle_{weak}} \textbf{S}_{i} \cdot \textbf{S}_{j} + J_{2} \sum_{\langle i,j\rangle_{strong}} \textbf{S}_{i} \cdot \textbf{S}_{j},
\label{eq:J1-J2}
\end{equation}
where $J_{1}$, $J_{2} > 0$ are antiferromagnetic couplings on nearest-neighbor bonds, with $J_1$ ($J_2$) denoting the weak (strong) bonds as illustrated in Fig.~\ref{fig:model} (a). All couplings act between nearest-neighbor sites; the distinction between $J_1$ and $J_2$ refers to the bond strength, not the bond range. In the Néel-ordered phase that appears for large $J_{1}/J_{2}$, the continuous $SU(2)$ spin-rotation symmetry is spontaneously broken, resulting in 
$N_{G} = 2$ gapless Goldstone modes. Upon tuning the control parameter $J_{1}$, the system undergoes a quantum phase transition at a critical value $J_{1,c} \approx 0.52337$ (fixing $J_{2}=1$)\cite{Wang2025Universal}. This transition is continuous and belongs to the (2+1)-dimensional $O(3)$ universality class, with a correlation-length critical exponent $\nu = 0.7073$\cite{Guillou1980Critical,Guillou1989Accurate,Rosenstein1993Critical,Guida1998Critical}.

Separately, we introduce the bilayer Heisenberg-Ising model on a square lattice shown in Fig.~\ref{fig:model} (b), whose Hamiltonian is:
\begin{equation}
    H= -J_{z} \sum_{\alpha=1,2}\sum_{\langle i,j\rangle} S^{z}_{i,\alpha}S^{z}_{j,\alpha} + J \sum_{i} \textbf{S}_{i,1} \cdot \textbf{S}_{i,2},
\label{eq:bilayer}
\end{equation}
here $S^z_{i,\alpha}$ is the $z$-component of a spin on layer $\alpha$, 
$J_{z} > 0$ is a ferromagnetic Ising coupling within each layer, and 
$J>0$ is an antiferromagnetic Heisenberg interlayer coupling. For sufficiently large $J_{z}$ the ground state develops ferromagnetic Ising order along the $z$-direction, which spontaneously breaks the discrete $Z_{2}$ symmetry of $S^{z}$. For the $1 \ll J_{z}/J$, the ground state has an intralayer-FM-interlayer-AF order, which breaks the $Z_2$ spin flip symmetry. For the $J_{z}/J \ll 1$, the ground state approaches a direct product of inter-layer spin-singlet states, which has an energy gap. When the inter-layer coupling is fixed at $J=3.045$, the model exhibits an order-disorder phase transition as the intralayer coupling $J_{z}$ is varied, with the critical point located at $J_{z}=1.0$\cite{Wu2023Classical}. Tuning the intra-layer coupling $J_{z}$ drives a phase transition whose critical behavior falls into the Ising universality class, with a correlation-length critical exponent $\nu = 0.63012$\cite{Kleinert1999Critical,Pelissetto2002Critical,Deng2003Simultaneous,Wu2023Classical}.

Concerning the Shannon-R\'enyi entropy, in a phase with spontaneously broken continuous symmetry its scaling obeys a volume law supplemented by a subleading logarithmic correction that counts the number of Goldstone modes. 
\begin{equation}
    S_{q} = a_{q} N + l_{q} \ln{N} + b_{q},
\label{eq:volumlaw}
\end{equation}
where $a_{q}$ is a nonuniversal coefficient of the leading volume-law term, $N$ is the number of sites in the system, $l_{q}$ denotes the coefficient of the subleading logarithmic correction, and $b_{q}$ is a constant. For example, in a phase with spontaneously broken continuous symmetry, $l_{q}=\frac{q}{q-1} \frac{N_{G}}{4}$, where $N_{G}$ is the number of Goldstone modes~\cite{Luitz2014Shannon}. 

\section{Results}
\subsection{SR entropy}
\begin{figure}[htp]
\centering
\includegraphics[width=1.0\columnwidth]{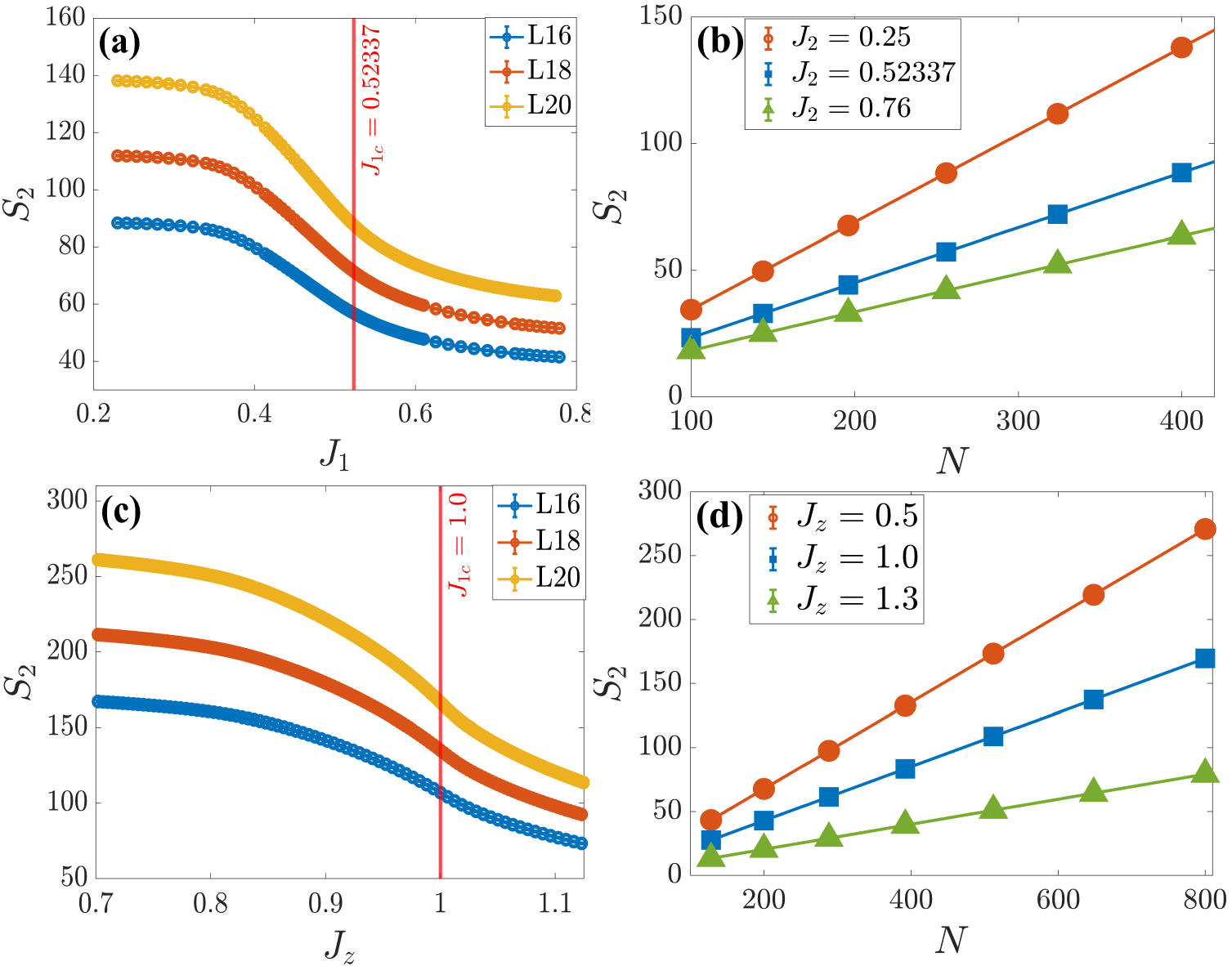}
\caption{Calculated SR entropy for the columnar dimerized Heisenberg model and the bilayer Heisenberg-Ising model with interlayer Heisenberg interaction. (a) SR entropy as a function of the system parameter $J_{1}$ for the columnar dimerized model. (b) Volume-law scaling of the SR entropy at $J_{1}=0.25$, $0.52337$, $0.76$. (c) SR entropy as a function of the system parameter $J_{z}$ for the bilayer Heisenberg-Ising model. (d) Volume-law scaling of the SR entropy at $J_{z}=0.5$, $1.0$, $1.3$. In panels (b) and (d), the numerical data are fitted according to Eq.~\eqref{eq:volumlaw}: in panel (b), the red, blue, and green lines correspond to the fits at $J_{1}=0.25$ [$l_{2}=0.00(9)$, $b_{2}=0.0(3)$], $J_{1}=0.52337$ [$l_{2}=0.02(4)$, $b_{2}=0.9(1)$], and $J_{1}=0.76$ [$l_{2}=1.025(9)$, $b_{2}=-1.05(4)$]; in panel (d), the red, blue, and green lines correspond to the fits at $J_{z}=0.5$ [$l_{2}=0.01(6)$, $b_{2}=0.0(2)$], $J_{z}=1.0$ [$l_{2}=0.04(2)$, $b_{2}=0.4(1)$], and $J_{z}=1.3$ [$l_{2}=0.01(1)$, $b_{2}=0.74(3)$].}
\label{fig:SRE}
\end{figure}

In this section, we show how the information on quantum phases and criticality can be extracted from the SR entropy. Unless otherwise stated, all results presented below are obtained in the $S^{z}$ basis, and the inverse temperature is scaled with the linear system size as $\beta=2L$ to ensure ground-state convergence. The coefficient of the subleading term of the SR entropy is universal~\cite{Nicolas2016Quantum}. Fig.~\ref{fig:SRE}~(a) presents the SR entropy at $q=2$ of the columnar dimerized Heisenberg model in Eq.~\eqref{eq:J1-J2} on the square lattice as a function of the system parameter $J_{1}$ (with $J_{2}=1$ fixed), where different colors denote different system sizes. The SR entropy remains smooth across the transition and exhibits no discontinuity at the critical point.

For the columnar dimerized model, the quantum critical point is located at $J_{1,c}=0.52337$. The transition is continuous and belongs to the $(2+1)$-dimensional $O(3)$ universality class. At the critical point, fitting Eq.~\eqref{eq:volumlaw} with $q=2$ to the data yields $l_{2}=0.02(4)$ and the $O(3)$-universality-class coefficient $b_{2}=0.9(1)$, which is consistent with existing numerical results \cite{Luitz2014Shannon,Luitz2017Quantum}. This constant is universal for a fixed point, similar to critical exponents.

On the two sides of the transition, the subleading scaling behavior differs markedly. For $J_{1,c} >J_{1}=0.25$, the system resides in the disordered phase, where no continuous symmetry is broken, leaving $N_{G}=0$ Goldstone modes; the fit gives $l_{2}=0$ and $b_{2}=0$ within statistical error, consistent with theoretical expectations~\cite{Luitz2014Shannon,Misguich2017Finite}. For $J_{1,c}<J_{1}=0.76$, the system enters the AF phase with spontaneously broken continuous symmetry, hosting $N_{G}=2$ Goldstone modes. In this phase the logarithmic subleading term extracted from the fit is $l_{2}=1.025(9)$, yielding $l_{\infty}=\frac{q-1}{q}l_{q}=0.513(5)$, which is reasonably close to the theoretical prediction $l_{\infty}=0.5$, with the residual deviation attributable to finite-size effects~\cite{Isaev2009Hierarchical,Luitz2014Universal}.

For the bilayer Heisenberg-Ising model, the quantum critical point is located at $J_{z,c}=1.0$ (with interlayer coupling $J=3.045$), and the transition belongs to the Ising universality class. At the critical point, fitting Eq.~\eqref{eq:volumlaw} yields $l_{2}=0.04(2)$ and $b_{2}=0.4(1)$, giving $l_{\infty}=\frac{q-1}{q}l_{q}=0.02(1)$ and $b_{\infty}=\frac{q-1}{q}b_{q}=0.20(5)$. In the disordered phase at $J_{z}=0.5$, we obtain $l_{2}=0$ and $b_{2}=0$ within statistical error, consistent with the theoretical prediction. In the $Z_{2}$ symmetry-breaking phase with $J_{z}=1.3$, the fitted parameters are $l_{2}=0.01(1)$ and $b_{2}=0.74(3)$; since $b_{q}$ is associated with the ground-state degeneracy, it is expected to be $b_{q>1}=\ln(2) \approx 0.6931$, which agrees with the fitted value within error bars. 

\subsection{Derivative of SR entropy}
\begin{figure}[htp]
\centering
\includegraphics[width=1.0\columnwidth]{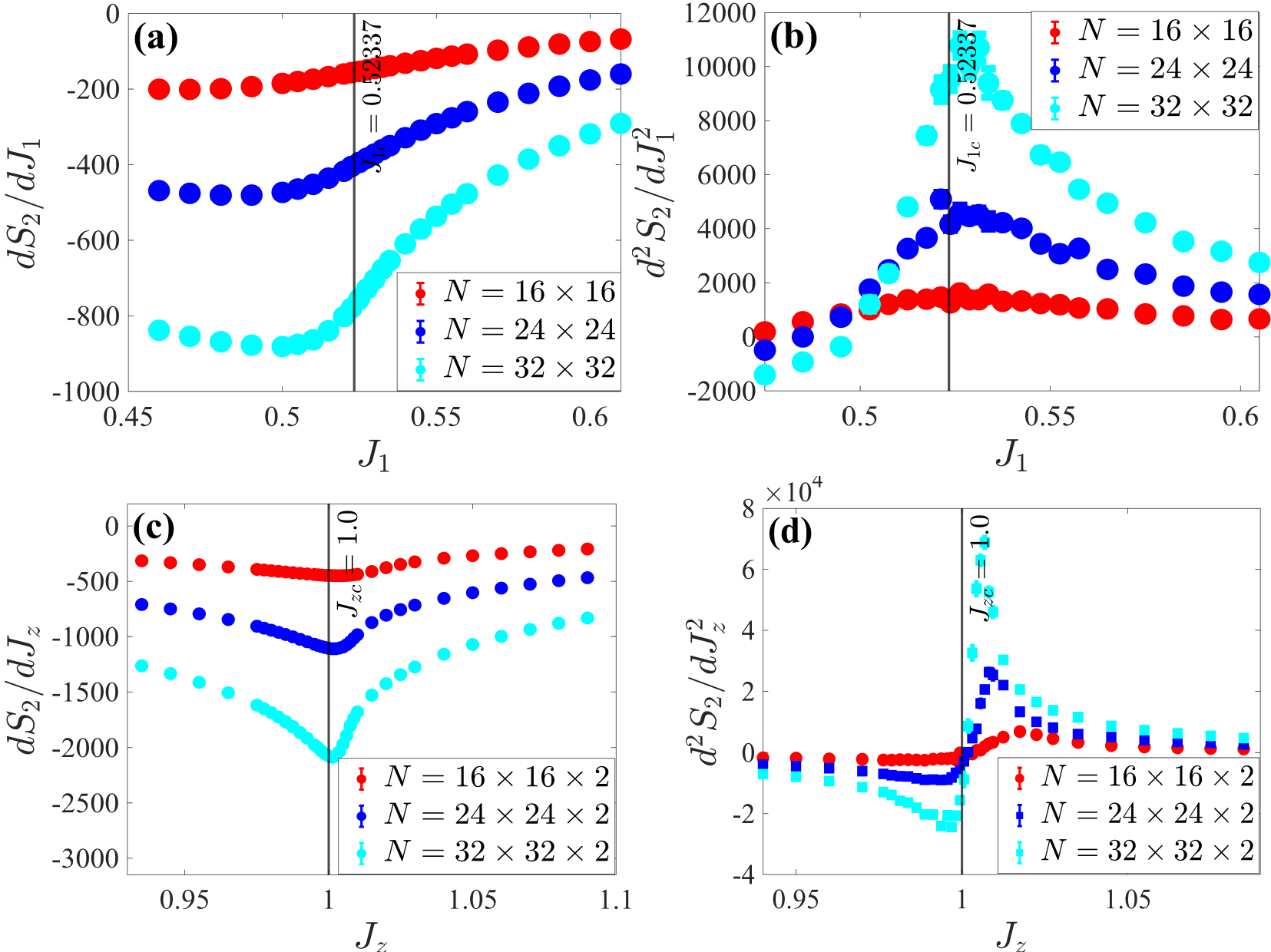}
\caption{First- and second-order derivatives of the SR entropy. Panels (a) and (b) show the first- and second-order derivatives of the SR entropy for the columnar dimerized Heisenberg model on a square lattice, plotted as a function of the parameter $J_{1}$. Panels (c) and (d) display the corresponding first- and second-order derivatives for the bilayer Heisenberg-Ising model with inter-layer Heisenberg interactions. }
\label{fig:derivative}
\end{figure}

In this section, we show that the scaling of the derivative of SR entropy can be collapsed to obtain the critical exponent $\nu$.
For the columnar dimerized Heisenberg model, the first- and second-order derivatives of the SR entropy with respect to the coupling $J_1$ in the vicinity of the transition point are shown in Figs.~\ref{fig:derivative}(a) and \ref{fig:derivative}(b), respectively. The first-order derivative varies smoothly across the transition and develops no peak; its slope at the transition point, however, steepens progressively with increasing system size. The second-order derivative develops a peak at the transition point whose height grows with increasing system size, indicating that the second-order derivative diverges in the thermodynamic limit. At any finite $L$ both derivatives are necessarily analytic, and the observed growth with system size is the finite-size precursor of the critical singularity at $J_{1,c}$.

For the Ising transition in the bilayer Heisenberg-Ising model, the first-order derivative of the SR entropy with respect to $J_z$ in the vicinity of the transition point is shown in Fig.~\ref{fig:derivative}(c), and the corresponding second-order derivative is given in Fig.~\ref{fig:derivative}(d). In this case, the first-order derivative itself develops a peak near the transition point whose height grows with system size, indicating that it diverges in the thermodynamic limit. Correspondingly, the second-order derivative changes sign from negative to positive across the transition, crossing zero near the peak position of the first-order derivative. The singular system-size dependence observed for both models is analyzed quantitatively through the data collapse presented below.

\begin{figure}[htp]
\centering
\includegraphics[width=1.0\columnwidth]{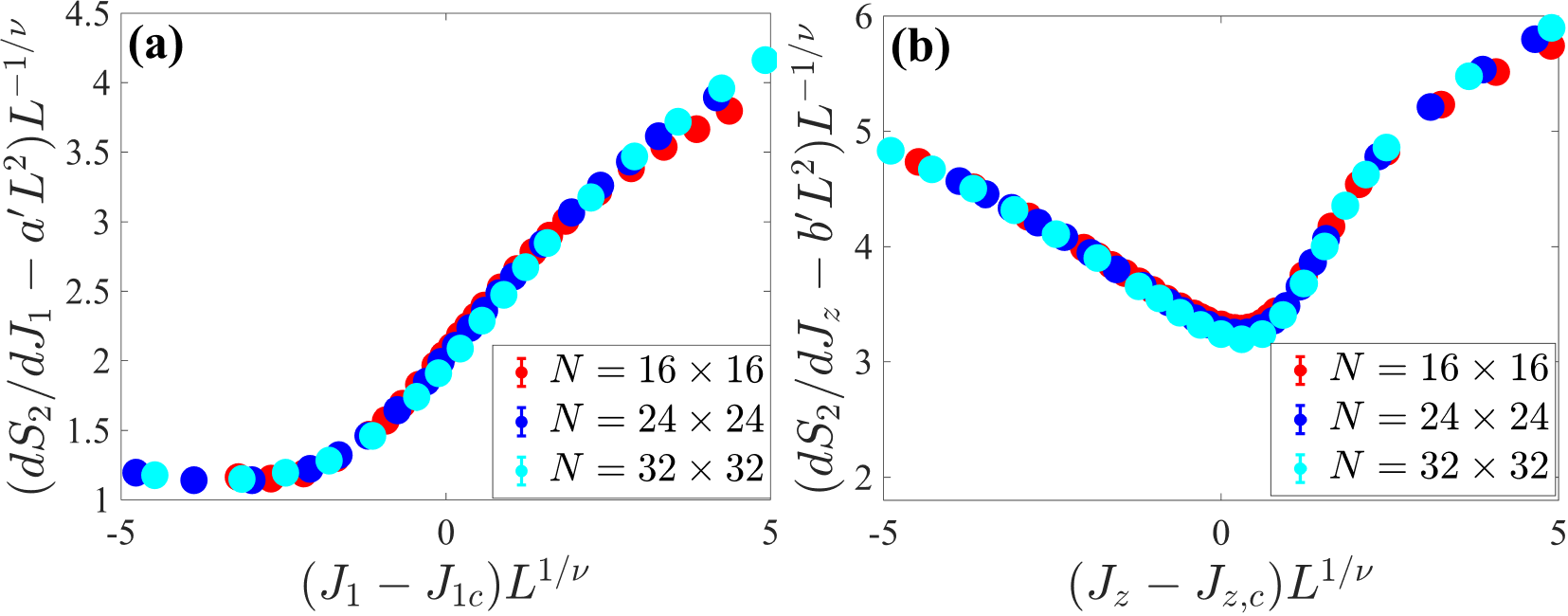}
\caption{Data collapse for the first derivatives of the SR entropy for both the columnar dimerized Heisenberg model on the
square lattice and the bilayer Heisenberg-Ising model with intralayer Ising coupling and interlayer Heisenberg coupling. (a) the x-axis is the scaled variable $(J_{1}-J_{1,c})L^{1/\nu}$, and the y-axis is the scaled variable $(d S_{2}/dJ_{1} - a^{\prime} L^{2}) L^{-1/\nu}$. (b) the x-axis is the scaled variable $(J_{z}-J_{z,c})L^{1/\nu}$, and the y-axis is the scaled variable $(d S_{2}/dJ_{z} - a^{\prime} L^{2}) L^{-1/\nu}$.}
\label{fig:dacoll}
\end{figure}

\subsection{Scaling theory for the SR entropy}
Previous studies have established that, at certain quantum critical points, the SR entropy contains a universal constant beyond its leading volume-law contribution
\cite{Stephan2009Shannon,Luitz2014Universal},
\begin{equation}
    S_q(g_c,L)=a(g_c)L^2+b_q+\cdots,
    \label{eq:scaling_SS2}
\end{equation}
where $a(g_c)$ is a nonuniversal coefficient and $b_q$ is a dimensionless constant associated with the critical state. We propose that this constant can be extended to a scaling function in the vicinity of the critical point, thereby connecting the SR entropy to the correlation length and the critical behavior of the system.

Our reasoning is the following. A universal contribution should be independent of microscopic lattice details and should instead be governed by long-wavelength physics. Near a continuous quantum phase transition, the characteristic length of this physics is the diverging critical length $\xi$. Since the entropy is dimensionless, the universal contribution should, under the finite-size scaling hypothesis, be described by a dimensionless function of the ratio $L/\xi$. We therefore promote the critical constant $b_q$ to a scaling function $S_q^0$ describing the evolution of this contribution near the transition.

According to the finite-size scaling hypothesis~\cite{fisher1972scaling, Sandvik2010Computational}, the scaling function should connect the critical limit to the nearby thermodynamic regime. At fixed small but nonzero distance from the critical point, the finite critical length should allow $S_q^0$ to contribute to the volume law. At the critical point, $\xi$ diverges and the finite system size provides the cutoff for the long-wavelength fluctuations. The contribution then remains of order unity and approaches the critical constant $b_q$. Therefore, in two spatial dimensions, the leading extensive behavior should be 
\begin{equation}
    S_q^0\sim r\frac{L^2}{\xi^2}.
\end{equation}
This provides a scaling description that connects the constant contribution at criticality to an extensive contribution in the nearby thermodynamic regime.

Using $\xi\propto|g-g_c|^{-\nu}$, the ratio $L/\xi$ can be expressed in terms of the conventional signed scaling variable $x=(g-g_c)L^{1/\nu}$, since $L/\xi\propto|x|^\nu$. Expressing the scaling function in terms of $x$, we obtain
\begin{equation}
    S_q(g,L)=a(g)L^2+S_q^0(x)+\cdots,
    \qquad x=(g-g_c)L^{1/\nu},
    \label{eq:scaling_S2}
\end{equation}
where $a(g)$ denotes the regular volume-law background and $S_q^0(0)=b_q$. The asymptotic behavior $S_q^0(x)\sim |x|^{2\nu}$ reproduces the singular extensive contribution proportional to $L^2/\xi^2$.

The scaling form of the derivative follows naturally by differentiating Eq.~\eqref{eq:scaling_S2} with respect to $g$:
\begin{equation}
    \frac{dS_q}{dg}
    =a'(g)L^2+S'_q(x)L^{1/\nu}+\cdots,
    \label{eq:dS2_scaling}
\end{equation}
where $S'_q(x)\equiv dS_q^0(x)/dx$ is a dimensionless scaling function. The factor $L^{1/\nu}$ follows from $dx/dg=L^{1/\nu}$ and describes the singular finite-size contribution associated with the diverging critical length.

Although this reasoning is straightforward, the proposed framework gives a direct, quantitative connection between the SR entropy, its parametric derivative, and quantum criticality. The universal constant becomes the critical value of a scaling function, while the singular response of the entropy is governed by the correlation-length exponent $\nu$. In the following, we examine this proposal for $q=2$ and use the derivative scaling to extract $\nu$ near the $O(3)$ and Ising quantum critical points.

Fig.~\ref{fig:dacoll} presents the data collapse of the first-order derivatives of the SR entropy with respect to the system parameter in the vicinity of the phase transition point, for both the columnar dimerized Heisenberg model on the square lattice and the bilayer Heisenberg-Ising model with intralayer Ising coupling and interlayer Heisenberg coupling. Panel~(a) of Fig.~\ref{fig:dacoll} display the data collapse of the first-order derivatives of the SR entropy with respect to $J_{1}$ for the columnar dimerized model, obtained at system sizes $L = 16, 24, 32$. Panel~(b) show the corresponding data collapse of the first-order derivatives with respect to $J_{z}$ for the bilayer Heisenberg-Ising model at the same set of system sizes. According to the scaling form Eq.~\eqref{eq:dS2_scaling}, the horizontal axis is uniformly rescaled as $(g - g_{c}) L^{1/\nu}$, while the vertical axis is rescaled as $\bigl(d S_{2}/d g - a^{\prime} L^{2}\bigr) L^{-1/\nu}$ for the first-order derivative at $q=2$. In the data collapse, the correlation-length exponent $\nu$ is treated as a free fitting parameter determined by optimizing the collapse quality. For the columnar dimerized model, whose transition belongs to the (2+1)-dimensional O(3) universality class, the optimal collapse yields $\nu = 0.701(5)$, in good agreement with the best available estimate $\nu = 0.7073$\cite{Guillou1980Critical,Guillou1989Accurate,Rosenstein1993Critical,Guida1998Critical}. For the bilayer Heisenberg-Ising model, whose transition belongs to the (2+1)-dimensional Ising universality class, equivalently the three-dimensional classical Ising universality class, the optimal collapse yields $\nu = 0.63(1)$, consistent with the best available estimate $\nu = 0.63012$~\cite{Kleinert1999Critical,Pelissetto2002Critical,Deng2003Simultaneous,Wu2023Classical}. The excellent data collapse, together with the fitted exponents agreeing with the known values, demonstrates that the scaling form Eq.~\eqref{eq:dS2_scaling} proposed in this work provides a quantitative description of the Shannon-R\'enyi entropy in the vicinity of the phase transition point.

\section{Conclusion}
In this work, we have established a field-theoretic spacetime manifold representation for the Shannon-R\'enyi entropy, which had remained elusive despite the mature manifold formulations for entanglement entropy and negativity. By mapping the $q$-th SR entropy onto $q$ replica loops glued at the imaginary-time boundaries, we recast it as a ratio of generalized partition functions, $S_q = \frac{1}{1-q}\ln[Z_G^{(q)}/Z^q]$, bringing it into a form directly amenable to large-scale quantum Monte Carlo simulation. This manifold construction is the conceptual cornerstone of the present work: it transforms the SR entropy from a quantity accessible only through brute-force histogramming into one that admits a polynomial-complexity, numerically exact evaluation. Just as the sheeted manifold representation of entanglement entropy unlocked QMC studies of R\'enyi entanglement, the glued-loop manifold introduced here unlocks QMC studies of the SR entropy.

Building on this representation, we developed a unified reweight-annealing algorithmic framework with two complementary strategies. The virtual-parameter scheme interpolates between the unglued ($\lambda=0$) and fully glued ($\lambda=1$) limits via a binomial-weighted mixture over partial gluing configurations, with the target ratio expressed as a telescoping product of strongly overlapping adjacent distributions. The real-parameter scheme replaces the artificial interpolation by a physical coupling constant, so that every intermediate distribution corresponds to a physically meaningful Hamiltonian and the reweighting factors reduce to simple moments of SSE operator counts. Both strategies eliminate the exponential bottleneck that has long restricted QMC studies of the SR entropy, reducing the exponential computational cost to polynomial. We benchmarked this framework on two paradigmatic models: the spin-$\frac{1}{2}$ columnar dimerized Heisenberg model and the bilayer Heisenberg-Ising model. In the former, the subleading logarithmic coefficient of the SR entropy in the antiferromagnetic phase yields the Goldstone-mode count $N_G=2$; in the latter, the subleading constant in the $Z_2$-symmetry-breaking phase reproduces the ground-state degeneracy $\ln 2$. These benchmarks confirm that the SR entropy, computed within our framework, serves as a quantitative probe of symmetry-breaking patterns and low-energy excitation spectra.

Most importantly, we have for the first time established a scaling theory for the SR entropy in the vicinity of a quantum critical point and demonstrated that universal critical information can be extracted from it. We proposed the scaling form
\begin{equation}
    S_q = a(g)L^2 + S_q^0\left((g-g_c)L^{1/\nu}\right),
\end{equation}
which requires neither a spatial bipartition nor a partial trace. Differentiating this form yields the scaling of the first derivative, $dS_q/dg = a'(g)L^2 + (dS_q^0/dx)L^{1/\nu}$, whose data collapse directly yields the correlation-length exponent $\nu$. We verified this scaling theory on two paradigmatic models belonging to distinct universality classes, namely the $(2+1)$-dimensional $O(3)$ transition and the Ising transition, and obtained excellent collapse for both. This establishes the SR entropy as a standalone, order-parameter-free diagnostic of quantum criticality, capable of extracting the correlation-length exponent from a single family of finite-size data.

Finally, because the SR entropy is defined in a fixed computational basis and is accessible through single-shot snapshots in cold-atom and Rydberg-array experiments, our framework closes the loop between theory and quantum simulation: it provides a scalable, numerically exact method for computing a quantity that is directly measurable in state-of-the-art quantum simulators. This opens the door to quantitative benchmarking of quantum simulator outputs against theoretical predictions, and to the systematic extraction of critical exponents from experimental snapshot data. More broadly, the manifold representation and reweight-annealing framework developed here are not limited to the spin models studied in this work; they can be generalized to fermionic and bosonic systems, to higher-order R\'enyi indices, and to other information-theoretic quantities that share the same diagonal structure. We anticipate that this work will catalyze a new generation of QMC studies of quantum-information observables in strongly correlated systems.

\section{Acknowledgement} This project is supported by the Scientific
Research Project (No. WU2025B011), Feng-Ying Career Development Chair Fund and the Start-up Funding of Westlake
University, and the Natural Science Foundation of China (Grant No. 12674188), and Quantum MetaKey (QuaMeK) Computing. The authors thank the IT service office and the high-performance
computing center of Westlake University.
\bibliographystyle{apsrev4-1}
\bibliography{bibliography}
\clearpage
\end{document}